\documentclass[useAMS,usenatbib]{mn2e}
 \usepackage[dvips]{graphicx}
 \usepackage[english]{babel}

\title[The spin of late-type galaxies at redshifts $z\le 1.2$]
{The spin of late-type galaxies at redshifts $\bmath{z\le 1.2}$}
\author[Cervantes-Sodi, Hernandez, Hwang, Park \& Le Borgne]
{Bernardo~Cervantes-Sodi$^{1}$\thanks{E-mail: bernardo@shao.ac.cn},
X.~Hernandez$^{2}$,
Ho~Seong~Hwang$^{3}$,
Changbom~Park$^{4}$
\newauthor
and Damien~Le~Borgne$^{5}$
\\
$^{1}$Partner Group of the Max Planck Institute for Astrophysics
and Key Laboratory for Research in Galaxies and Cosmology\\
of Chinese Academy of Sciences, Shanghai Astronomical Observatory,
Nandan Road 80, Shanghai 200030, China
 \\
$^{2}$Instituto de Astronom\'\i a,
Universidad Nacional Aut\'onoma de M\'exico
A. P. 70--264,  M\'exico 04510 D.F., M\'exico \\
$^{3}$Smithsonian Astrophysical Observatory, 60 Garden Street, Cambridge, MA 02138, USA
\\
$^{4}$Korea Institute for Advanced Study, Dongdaemun-gu, Seoul 130-722, Korea\\
$^{5}$Institut d'Astrophysique de Paris, UMR 7095, CNRS, UPMC Univ. Paris 06, 98bis boulevard Arago, F-75014 Paris, France
\\
}
\begin{document}

\date{In original form 2012 April 19}

\pagerange{\pageref{firstpage}--\pageref{lastpage}} \pubyear{2012}

\maketitle

\label{firstpage}

\begin{abstract}

We study the evolution of the galactic spin using data of high redshift galaxies
in the fields of the Great Observatories Origins Deep Survey (GOODS).
Through simple dynamical considerations we estimate the spin for the
disc galaxies in our sample and find that its distribution is consistent
with that found for nearby galaxies. Defining a dimensionless angular
momentum parameter for the disc component of the galaxies ($\lambda_{d}$),
we do not find signs of evolution in the redshift range $0.4  \leq z \leq 1.2$.
We find that the mass and environmental dependencies of the spin of our
high redshift galaxies are similar to those of low-$z$ galaxies;
showing a strong dependence on mass, in the sense that low-mass systems present
higher $\lambda_{d}$ values than high-mass galaxies, with no significant dependence
on the environmental density. These results lead us to conclude that, although
individual disc galaxies might occasionally suffer strong evolution,
they evolve in such a way that the overall spin distribution of
the galactic population remains constant from $z\sim1$ to the present epoch.

\end{abstract}

\begin{keywords}
galaxies: evolution -- galaxies: fundamental parameters -- galaxies: general -- galaxies: statistics -- galaxies: structure
 -- galaxies: high-redshift.
\end{keywords}

\section{Introduction}

In the standard picture of galaxy formation, galaxy discs form out of
gas that slowly cools out of a hot gaseous halo, conserving its specific angular
momentum, and forms a disc at the centre of the potential well of a dark
matter halo
(White \& Rees 1978;
Fall \& Efstathiou 1980). The specific angular momentum of the pre-collapse
gas is generally assumed to be equal to that of
the dark matter, which is acquired by tidal torques in the early
Universe (Peebles 1969). This simple picture leads to predictions of
present day disc-galaxies that show reasonably good agreement with
observations (Dalcanton, Spergel \&
Summers 1997; Avila-Reese, Firmani \& Hernandez (1998);
Jimenez et al. 1998; Mo, Mao \& White 1998; de Jong \& Lacey 2000;
Pizagno et al. 2005; Dutton et al. 2007), however; later stages of
galaxy evolution must certainly include a large variety of processes
such as mergers (Naab, Jesseit \& Burkert 2006; Martig et al. 2012),
galaxy-galaxy interactions (Cervantes-Sodi, Hernandez \& Park 2010;
Hwang et al. 2011; Lee et al. 2012)and accretion of cold
gas through thin dense filaments (Kere{\v s} et al. 2005; Powell et al. 2011),
that shape the morphology of galaxies. 

If this complex scenario is actually taking place, it is remarkable
to find the presence of fully formed disc galaxies at high redshift
that highly resemble present day systems (Genzel et al. 2006), requiring
only moderate evolution after a rapid early assembly to reproduce the
nearby population of Milky Way-type galaxies. Observations of disc galaxies
at $z \leq 1$ show the existence of large discs, suggesting that
these kind of systems must be assembled prior to this epoch (Lilly et al. 1998;
Trujillo \& Aguerri 2004; Sargent et al. 2007; Kanwar et al. 2008). Not
only do these galaxies have similar sizes to their present day relatives, but
they also follow the same scaling relations with none or only mild evolution.
Barden et al. (2005) found weak or no evolution in the stellar-mass$-$size
relation for disc dominated galaxies back to $z = 1$. This result is consistent
with a passively evolving stellar population at a given mass, with no growth of
galaxy disks, or with the idea that in fact the galaxies are growing, but 
in such a way as to
evolve along the same stellar-mass$-$size relation. Recently, results
pointing in the same direction were reported by Ichikawa, Kajisawa
\& Akhlaghi (2012) with a sample extending up to $ z \sim 3$.

Similar conclusions are reached by studying other relations,
such as the Tully-Fisher (TF) that evolves, specially for the case of blue bands
that are highly sensitive to the recent star formation,
but presents only mild or no evolution for near infrared bands (Fern\'andez
Lorenzo et al. 2010) and no evolution at all for the case of the stellar mass
TF relation (Miller et al. 2011). High redshift galaxies, up to $z = 3.5$ do
not show evidence of evolution in the fundamental plane
defined by star formation rate, metallicity and stellar
mass (Lara-L\'opez et al. 2010). In addition, a
roughly constant number density of large disc galaxies since $ z \sim 1$ (Lilly
et al. 1998; Sargent et al. 2007), and a mass function that does not present strong
features of evolution in the same redshift range (Brinchmann \& Ellis 2000), all 
indicate that most of the large disc galaxies have experienced little evolution 
in the last eight billion years.

Along with mass, angular momentum shapes fundamental properties of galaxies,
and plays a preponderant role in establishing fundamental relations, such as the
TF (Koda, Sofue \& Wada 2000). In previous studies{\bf ,} we have shown using different
samples of galaxies, how the overall morphology of disc galaxies is intimately
linked to the galactic angular momentum (Hernandez \& Cervantes-Sodi 2006;
Cervantes-Sodi \& Hernandez 2009; Cervantes-Sodi et al. 2011b),
and making use of extended samples from the Sloan Digital Sky Survey
(SDSS, Choi et al. 2010), we have also looked for dependences of the spin
parameter on the total mass of the galaxies (Cervantes-Sodi et al. 2008) and on
local and global environment (Cervantes-Sodi et al. 2008; Cervantes-Sodi,
Hernandez \& Park 2010; Cervantes-Sodi et al. 2011a). The aim of this work is to
get empirical distributions of the galactic spin for a sample of high redshift
galaxies, to look for any evolution on the distributions of this parameter, and 
compare mass and environment dependencies of the high redshift sample with 
previous results for local galaxies. This paper is organised as follows: 
Section 2 gives a brief review of our model to estimate the spin for disc 
galaxies in the sample, the sample details are described in Section 3. In 
Section 4 we present the general results, with general conclusions appearing 
in Section 5.
Throughout, we adopt $h=0.7$ and a flat $\Lambda$CDM cosmology 
  with density parameters 
  $\Omega_{\Lambda,0}=0.73$ and $\Omega_{m,0}=0.27$.

\section{Estimation of the spin from observable parameters}

A traditional way to characterise the galactic angular momentum is through the
$\lambda$ spin parameter, as defined by Peebles (1971);

\begin{equation}
\label{Lamdef}
\lambda = \frac{L \mid E \mid^{1/2}}{G M^{5/2}},
\end{equation}

where $E$, $M$ and $L$ are the total energy, mass and angular momentum of the configuration, 
respectively. In 
Hernandez \& Cervantes-Sodi (2006), we derived a simple estimate of total $\lambda$ for dark halos
hosting disc galaxies 
in terms of observational parameters, based on two simple hypothesis: that the specific angular
momentum of dark matter and baryons are equal, and a constant small baryonic fraction, for systems
where the total energy and angular momentum are dominated by the dark matter component.
But serious questions arise about these hypothesis when studying galaxies at high redshift.
Regarding a constant baryonic fraction, a strong dependence on redshift has been suggested
(Moster et al.  2010; Behroozi, Conroy \& Wechsler 2010; Faucher-Gigu\`ere, Kere\v{s} \& Ma 2011),
although no clear consensus has been reached, specially for $z\geq 1$. Notice however, that the
lack of evolution of the baryonic Tully-Fisher relation argues for little evolution of the baryonic
galactic fraction.
Concerning the first hypothesis, using high-resolution cosmological simulations,
several authors claim that the equality at all times for an evolving galaxy is
unrealistic (e. g. Dutton \& van den Bosch 2011), but no general consensus has
been reached. Recent studies report a large variety of results, some showing a
clear one-to-one correlation between the specific angular momentum of
baryons and dark matter (Zavala, Okamoto \& Frenk 2008), while others
present analytic expressions in terms of the dark matter fraction
(Sales et al. 2009) or a combination of virial mass and redshift
(Kimm et al. 2011), to establish a relation between the angular momentum
of both components.

To avoid complications, we must therefore use 
an angular momentum parameter which focuses on the dynamics of the stellar disc.
We retain the quantitative and objective nature of the study, 
and account for the angular momentum focusing only on the stellar component to define a
disc dimensionless angular momentum parameter $\lambda_{d}$ as we did in Cervantes-Sodi
et al. (2011). Here we give a brief account of the model.
We consider a disc for the stellar component of the galaxy with
an exponential surface mass density $\Sigma(r)$;

\begin{equation}
\label{Expprof}
\Sigma(r)=\Sigma_{0} e^{-r/R_{d}},
\end{equation} 

where $r$ is a radial coordinate and $\Sigma_{0}$ and $R_{d}$ are two constants which are allowed 
to vary from galaxy to galaxy, and assume the presence of a dark matter halo responsible for
establishing a rigorously flat rotation curve $V_{d}$ throughout the disc.

From equation~\ref{Expprof}, the total disc mass is

\begin{equation}
\label{Discmass}
M_{d}=2 \pi \Sigma_{0} R_{d}^{2}.
\end{equation}

This, combined with our flat rotational curve, leads to an angular momentum of
$L_{d}=2V_{d}R_{d}M_{d}$.
Assuming the disc to be a virialized dynamical system, the total energy
can be obtained from the total kinetic energy, estimated as arising merely
from the dominant rotation. In this case, the kinetic energy
of the disc is $T_{d}=M_{d}V_{d}^{2}/2$.

These assumptions allow us to express $\lambda_{d}$ as

\begin{equation}
\label{Ld}
\lambda_{d}= \frac{L_{d} \mid T_{d} \mid^{1/2}}{G M_{d}^{5/2}} = \frac{2^{1/2} V_{d}^{2} R_{d}}{G M_{d}}.
\end{equation}

Finally, we introduce a stellar TF relation
(Miller et al. 2011): 
$M_{d}=A_{TF} V_{d}^{3.869}$, to replace the dependence on $V_{d}$ for a dependence
on the stellar mass available in our sample (see Section 3), to obtain our final estimation of $\lambda_{d}$.

Note that equation~\ref{Ld} is the same expression we derived in Hernandez \& Cervantes-Sodi
(2006) to estimate the traditional $\lambda$ spin parameter for dark matter haloes hosting
disc galaxies, divided by the stellar fraction of the galaxy; this because to obtain the total
spin parameter we were assuming angular momentum conservation for both components, and a
constant baryonic fraction, linking the specific angular momentum and mass of the dark
matter to those quantities of the disc component. In our current work however, we do
not attempt to constrain the physical characteristics of the halo, but just
consider its participation in establishing the flat rotation curve throughout the disc.
The parameter $\lambda_{d}$ is hence not a $\lambda$ parameter in the sense of the definition
of equation~\ref{Lamdef}, but merely an estimate of a dimensionless angular momentum for a galactic disc,
expected to correlate tightly with all type-defining properties.

\section{GOODS sample}

The sample of high redshift galaxies used for this study is an updated version
of the one presented in Hwang \& Park (2009). Here, we give a brief description
of the sample, and we refer the reader to Hwang et al. (2011) for a detailed description
of the data.

We use a spectroscopic sample of galaxies from the Great Observatories Origins Deep Survey
(GOODS), which is a deep multiwavelength survey from NASA's Great Observatories,
Spitzer, Hubble, and Chandra, ESA's Herschel and XMM-Newton, and from the most
powerful ground-based facilities, with a total observing area approximately 320 arcmin$^2$ from two carefully selected regions centered
on the Hubble Deep Field North (GOODS-N), and
Chandra Deep Field South (GOODS-S).
From the vast spectroscopic data for GOODS sources in the
literature, we used a total number of 6958 galaxies whose spectroscopic redshifts
are reliable over the whole fields of GOODS-N (Cohen et al. 2000; Cowie et al. 2004; Wirth et al. 2004;
Reddy et al. 2006; Barger et al. 2008; Cooper et al. 2011a;)
and GOODS-S (Szokoly et al. 2004;
Le F\'evre et al. 2004; Mignoli et al. 2005; Vanzella et al. 2005,
2006, 2008; Ravikumar et al. 2007; Popesso et al. 2009;
Kurk et al. 2009; Balestra et al. 2010; Silverman et al. 2010;
Xia et al. 2011; Cooper et al. 2011b), respectively, with typical error
of $4 \times 10^{-4}$.
In our analysis, a volume-limited sample with
$M_r \leq -20.0$ and $0.4\leq z\leq1.2$ is used.
The rest frame $r$-band absolute magnitude $M_r$ of galaxies was computed based
on the ACS plus near-infrared (NIR) photometry with $K$-corrections (Blanton \& Roweis 2007).
The $1.1(z-0.1)$ term was added to $M_r$ for the evolution correction (Wolf et al. 2003).
Stellar masses are computed from $U$-band to IRAC 4.5 $\mu$m photometric data
using Z-PEG (Le Borgne \& Rocca-Volmerange 2002). The templates used for the stellar
mass estimates are determined from P{\'E}GASE.2 (Fioc \& Rocca-Volmerange 1999)
assuming a Salpeter initial mass function (Salpeter 1955).
The templates were produced using different scenarios
for the star formation history (see Le
Brogne \& Rocca-Volmerange 2002) varying the star-formation efficiency and
infall timescales, raging from a pure starburst to a continuous star
formation rate, lasting from 1 Myr to 13 Gyr with the requirement for the
templates to be younger than the age of the Universe at any redshift.
The models include dust computed from P{\'E}GASE.2, consistently with
the star formation histories. The typical amount of reddening by dust is
within the range $0.0 \leq E(B-V) \leq 0.15$. The maximum error
associated to the stellar mass estimate is 0.3 dex
(see Elbaz et al. 2011 for more details).

This estimate of stellar mass is the one used to infer $V_{d}$ needed in
equation~\ref{Ld} to calculate $\lambda_{d}$. Choosing a stellar
TF relation has a major advantage over traditional TF relations based on specific
bands. As recently reported by several authors (Kassin et al. 2007,
Fern{\'a}ndez Lorenzo et al. 2010,
Miller et al. 2011), the stellar TF seems not
to evolve with redshift, at least up to redshift $z \sim 1.7$
(Miller et al. 2012), and presents the smallest scatter among other TF relations,
with intrinsic scatter of 0.058 dex, which is comparable to that seen in local TF
relations (i.e., $\sim$0.049 in Pizagno et al. 2005).

Given that our spin estimate is suitable only for disc galaxies, we need to
segregate the galaxies in our sample according to their morphology. To do
that, we adopt the segregation criteria by Hwang \& Park (2009) and Hwang et al.
(2011), where galaxies are classified into early (E/S0) and late (S/Irr) types by visual
inspection. Early-type galaxies are those with little fluctuation in the surface
brightness and color and with good symmetry, while late-type
galaxies show internal structures and/or color variations in the
pseudocolor images. For late-type galaxies
we use as a proxy for the scale length $R_{d}=R_{e}/1.68$, where
$R_{e}$ is the half light radius in the $z$-band.

In Section 4 we will study the dependence of the spin on the environment.
To account for the large-scale environment we consider a surface
galaxy number density estimated from the five nearest
neighbour galaxies ($\Sigma_{5}$), defined by

\begin{equation}
\Sigma_{5} = 5 (\pi D^{2}_{p,5})^{-1}, 
\label{Sigma_5}
\end{equation}

where $D_{p,5}$ is the projected proper distance to the 
5th-nearest neighbour, which is identified among the neighbour galaxies with
$M_r \leq -19.5$ that have velocities relative to the target galaxy less than
1000 km s$^{-1}$, so as to exclude foreground and background galaxies.

We will also consider the distance to the nearest neighbour galaxy as a
small-scale environmental parameter. The nearest neighbour galaxy
of a target galaxy with absolute magnitude M$_{r}$ is the one with
the smallest projected separation distance on the sky to the galaxy,
is brighter than $M_{r} + \Delta M_{r}$
among those in the sample, with $\Delta M_{r}=0.5$ and has relative velocity
with respect to the target galaxy less than 
$\Delta \upsilon=|\upsilon_{\rm neighbors}- \upsilon_{\rm target}|=660$
km s$^{-1}$ for early-type target galaxies and less than
$\Delta \upsilon=440$ km s$^{-1}$ for late-type target galaxies. These
velocity difference limits are 10\% larger than those that we have
used for SDSS galaxies in previous studies, i.e. Fig 1 of Park et al. (2008),
because of the larger redshift uncertainties for GOODS galaxies as
compared with SDSS galaxies.

The virial radius of a galaxy within which 
  the mean mass density is 200 times the critical density 
  of the universe ($\rho_c$), is calculated by
\begin{equation}
r_{\rm vir}=(3 \gamma L/4\pi)^{1/3} (200\rho_c)^{-1/3},
\label{Rvir1}
\end{equation}
where $L$ is the galaxy luminosity, and $\gamma$ the mass-to-light ratio.
Here, the mass associated with a galaxy plus dark halo
  system is assumed to be proportional to the $r$-band luminosity of the galaxy.
We assume that the mass-to-light ratio of early-type galaxies
  is on average twice as large as that of late-type galaxies
  at the same absolute magnitude $M_r$,
  which means $\gamma$(early)$=2\gamma$(late) 
 following Choi et al. (2007) for SDSS galaxies and Hwang \& Park (2009)
 for GOODS galaxies.
The critical density of the universe ${\rho}_c$ 
  is a function of redshift $z$
  [i.e. ${\rho}_c=3H^2(z)/(8\pi G)$] and 
 $\Omega_m(z)= \rho_b(z)/\rho_c (z)=\overline{\rho}(1+z)^3/\rho_c(z)$,
 where $\rho_b$ and $\overline{\rho}$ are the mean matter densities 
 in proper and comoving spaces, 
 respectively.
The Hubble parameter at $z$ is
  $H^2(z)=H^2_0 [\Omega_{m,0}(1+z)^3 +\Omega_{k,0}(1+z)^2+\Omega_{\Lambda,0}]$,
  where $\Omega_{m,0}$, $\Omega_{k,0}$, and $\Omega_{\Lambda,0}$ 
  are the dimensionless density parameters at the present epoch (Peebles 1993).
Then, the virial radius of a galaxy at redshift $z$ in proper space 
  can be rewritten by
  
\begin{equation}
r_{\rm vir} (z) = [3 \gamma L \Omega_{m,0} / (800\pi \overline{\rho}) / 
 \{ \Omega_{m,0}(1+z)^3 + \Omega_{k,0}(1+z)^2 + \Omega_{\Lambda,0} \} ]^{1/3}.
\label{Rvir}
\end{equation}

The mean mass density $\overline{\rho}$ was computed using
  the galaxies at $z=0.4-1.2$ with
  various absolute magnitude limits varying from $M_r= -16$ to $-20$.
Hwang et al. (2011) found that the mean mass density
  appears to converge when the magnitude cut is fainter than $M_r=-17.5$,
  which means that the contribution of faint galaxies
  is not significant because of their small masses.
In this calculation,
  each galaxy is weighted by the inverse of completeness
  according to its apparent magnitude and color (see Fig. 1 of Hwang \& Park 2009).
The final values are $\overline{\rho} =$ 0.017 and 0.013 $ (\gamma L)_{-20}$ (Mpc$^{-3}$)
  for GOODS-N and -S, respectively,
  where $(\gamma L)_{-20}$ is the mass of a late-type galaxy with $M_r=-20$.
According to our formula the virial radii of galaxies with
  $M_r=-20$ and $-21$ are 300  and 400 $h^{-1}$ kpc for early types,
  and 240 and 320 $h^{-1}$ kpc for late types, respectively.

The final volume limited sample contains 827 late-type galaxies
with accurate total stellar masses and reliable spectroscopic redshifts.

\section{Results}

Once having segregated the galaxies according to their morphology, and
with all the information required to apply equation~\ref{Ld}, we obtained
the $\lambda_{d}$ distribution for all the late type galaxies in our
sample. Usually, as is also the case here, this parameter is well described by a log-normal function of the form:

\begin{equation}
\label{Plam}
P(\lambda_{d0},\sigma_{\lambda_{d}};\lambda_{d}) d\lambda_{d}=
\frac{1}{\sigma_{\lambda_{d}}\sqrt{2\pi}}exp\left[-\frac{ln^{2}(\lambda_{d}/\lambda_{d0})}
{2\sigma_{\lambda_{d}}^{2}} \right] \frac{d\lambda_{d}}{\lambda_{d}}
\end{equation}

The parameters for the best-fitting distribution are $\lambda_{d0}=0.752$
and $\sigma_{\lambda_{d}}=0.499$. To compare directly with global
$\lambda$-distributions, we need to adopt a recipe to estimate the total
angular momentum and mass in terms of stellar parameters. To do so, we
first assume that the specific angular momentum of the disc and the halo
are equal (e.g., Mo et al. 1998; Zavala et al. 2008); and then we need to
choose a proper dark matter fraction. As in Hernandez et al. (2007),
for a constant dark matter fraction (F1), the $\lambda$-distribution for
this sample of high redshift galaxies is compatible with results of
low redshift galaxies,  with $\lambda_{0}=0.045$ and $\sigma_{\lambda}=
0.499$. The shape of the distribution strongly depends
on the assumed dark matter fraction;
just to illustrate this strong dependency we show in Fig 1 two other distributions;
one considering a dependence of the dark matter fraction on the stellar
surface density (F2) from Gnedin et al. (2007), 

\begin{figure}
\label{distributions}
\centering
\begin{tabular}{c}
\includegraphics[width=0.475\textwidth]{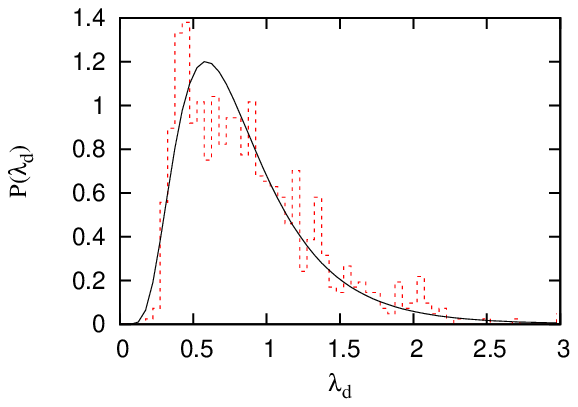} \\
\includegraphics[width=0.475\textwidth]{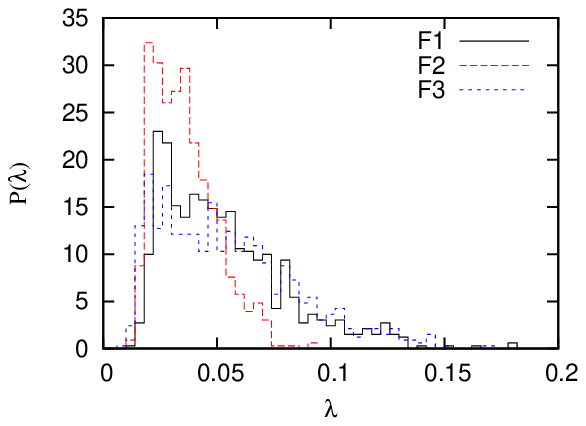}
\end{tabular}
\caption[ ]{\textit{Upper panel}: P($\lambda_{d}$) distribution for late-type
galaxies in the sample with the best lognormal fit to the data with parameters
$\lambda_{d0}=0.752$
and $\sigma_{\lambda_{d}}=0.499$. \textit{Lower panel}: P($\lambda$) distributions
considering three different functions for the dark matter fraction.}
\end{figure}

\begin{equation}
F = F_{0} \left( \frac{M_{*} R_{d}^{-2}}{10^{9.2} M_{\odot} kpc^{-2} }\right)^{p},
\end{equation}

where $p=0.2$, that produces a distribution with $\lambda_{0}=0.032$ and $\sigma_{\lambda}=0.392$; and following 
Faucher-Gigu\`ere, Kere\v{s} \& Ma (2011), one where the mass of the
dark matter halo is given in terms of the rotation velocity
and redshift:

\begin{equation}
M_{h}=10^{10} M_{\odot} \left(\frac{V_{d}}{50 km ^{-1}} \right) \left(\frac{1+z}{4} \right)
\end{equation}

resulting in $\lambda_{0}=0.046$ and $\sigma_{\lambda}=0.531$. 
As shown in Fig 1, the dependence on the assumed dark matter fraction is
strong, with clear changes of the spin distributions according to the
dark matter fraction chosen, but giving as a result $\lambda$-distributions
that are consistent with those previously found by different authors, as
summarized by Shaw et al. (2006), giving values in the range 
$0.03 < \lambda_{0}< 0.05$ and $0.48< \sigma_{\lambda}< 0.64$. To avoid the
complication of determining the dark matter fraction, plus the
systematic offsets due to the coefficients involved, and the possibility
that the specific angular momentum of dark matter and stars are not
equal (e. g. Kassin et al. 2012), we will present
the following results in terms of the disc spin $\lambda_{d}$, which can
be interpreted in terms of the global spin selecting a proper
dark matter fraction, assuming that the specific angular momentum of
dark matter can be traced by that of the baryons. 

\begin{figure}
\centering
\begin{tabular}{c}
\includegraphics[width=0.475\textwidth]{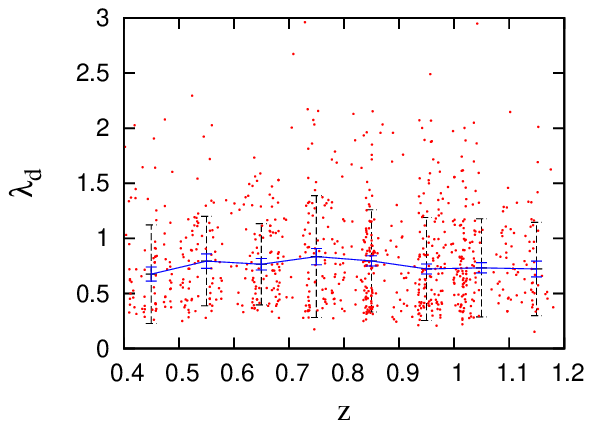}  \\
\includegraphics[width=0.475\textwidth]{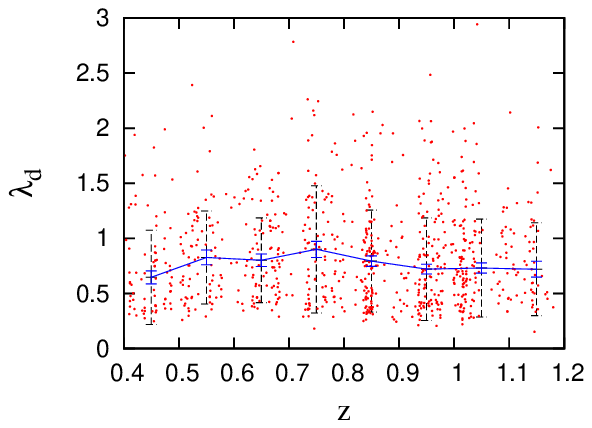}
\end{tabular}
\caption[ ]{\textit{Upper panel}:$\lambda_{d}$ value as a function of redshift, showing the median
$\lambda_{d}$ value for each bin, using a constant stellar mass-TF relation
for the whole sample. Error bars in solid blue lines denote the estimated
1$\sigma$ confidence intervals for the mean based on the bootstraping resampling method,
error bars in broken black lines denote the dispersion for each bin.
\textit{Lower panel}: Same as the upper panel but using three different stellar
mass-TF relations valid on three redshift ranges delimited on Table 1.}
\label{LdVsZ}
\end{figure}  

With our sample extending over a wide redshift range, we search for an
evolution with time. Fig. 2 shows $\lambda_{d}$
values as a function of redshift, with the sample divided into eight
bins, where the median $\lambda_{d}$ values are shown with error bars
in solid blue lines that represent the 1$\sigma$ confidence intervals based on the
bootstraping resampling method and error bars in broken black lines denote the dispersion for each bin; this convention will be followed
for the next figures. We clearly see that the typical 
values remain almost constant throughout the whole redshift range. 

Bearing in mind that we have chosen a specific TF relation
for all the galaxies in our sample, we check if the
lack of evolution we find for the disc spin comes directly form
the hypothesis of the validity of the same TF relation for the
galaxies in the redshift range $0.4\leq z\leq1.2$. Instead of using
one single stellar mass TF relation, we use three different relations
of the form 

\begin{equation}
log (M_{d}) = [a + b \times log (V_{d})] − log (M_{0}),
\end{equation}

where $M_{0}=10^{10}$ and the constants $a$ and $b$, taken from
Miller et al. (2011), are given in Table 1. The result of using
a combination of TF relations for the three different redshift
ranges is shown in Fig. 2 lower panel, where we can see that
even using these relations there is no evidence of evolution.

\begin{table}
\caption{Stellar mass TF relations.}
\begin{tabular}{|c|c|ccc|}

	\hline

    redshift range & $a$    &   $b$    \\
	\hline
	$0.2 \leq z \leq 1.2$ & 1.718  & 3.869 \\
	\hline
	$0.2 \leq z \leq 0.5$ & 1.755  & fixed \\
	$0.5 \leq z \leq 0.8$ & 1.684  & fixed \\
	$0.8 \leq z \leq 1.2$ & 1.720  & fixed \\	
	\hline
	
\end{tabular}
\end{table}

Given the lack of evolution of the $\lambda_{d}$ parameter on redshift
for the galaxies in our sample, a reasonable test is to check if these
galaxies present the same strong dependence of $\lambda_{d}$ on mass,
and a lack of dependence on environment, as present for
low redshift galaxies (Cervantes-Sodi et al. 2008; Berta et al. 2008;
Cervantes-Sodi et al. 2010).
The upper panel in Fig 3 shows $\lambda_{d}$ as a function of stellar mass,
showing a strong dependence, with high mass galaxies showing typically
low $\lambda_{d}$ values, as well as lower spread when compared with low
mass galaxies, just the same behaviour as the one showed by low redshift
galaxies (see Fig 5 in Cervantes-Sodi et al. 2008).

For samples of galaxies at low redshift, we previously reported a lack of
dependence of the spin parameter on the large-scale environment, which is in
good agreement with theoretical expectations (Lemson \& Kauffmann 1999); for
the present sample of high redshift galaxies we again recover the same result
as shown in Fig 3 lower panel.

\begin{figure}
\label{MassEnvironment}
\centering
\begin{tabular}{c}
\includegraphics[width=0.475\textwidth]{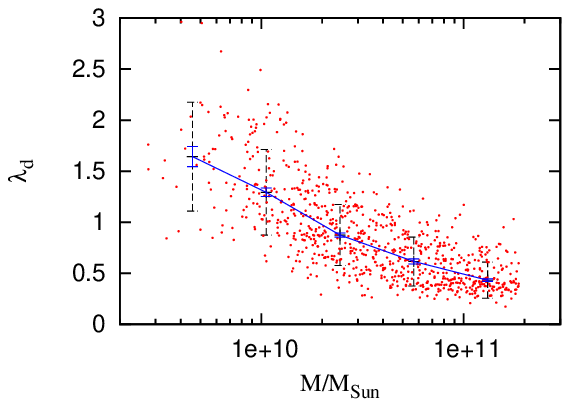} \\
\includegraphics[width=0.475\textwidth]{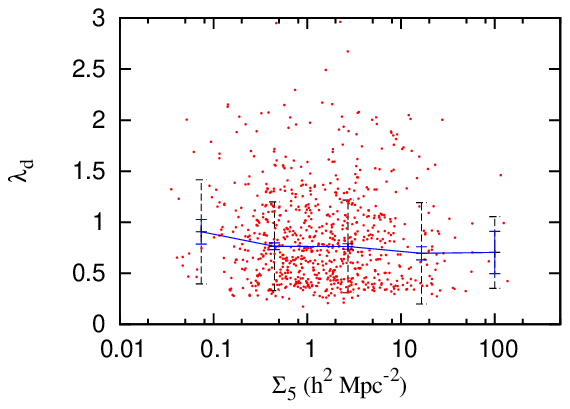}
\end{tabular}
\caption[ ]{Correlations between the disc spin parameter with stellar mass (\textit{upper panel}) and environmental density $\Sigma_{5}$(\textit{lower panel}).}
\end{figure}

\begin{figure}
\label{MassEnvironment}
\centering
\begin{tabular}{c}
\includegraphics[width=0.475\textwidth]{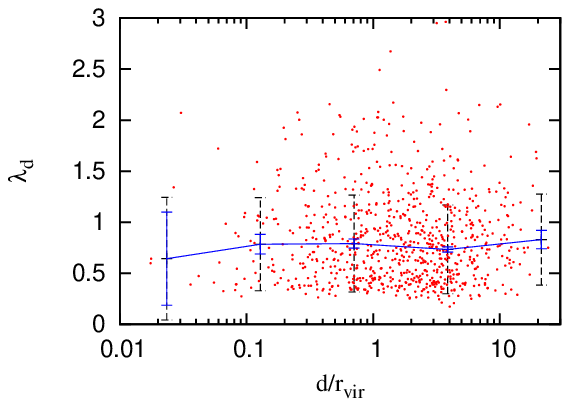} \\
\includegraphics[width=0.475\textwidth]{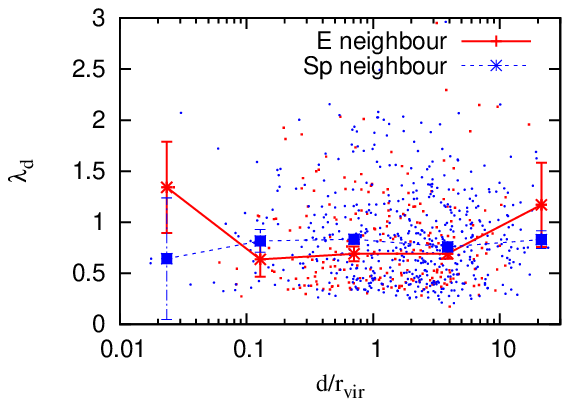}
\end{tabular}
\caption[ ]{\textit{Upper panel}:$\lambda_{d}$ as a function of the distance to the nearest
neighbour galaxy normalized by the virial radius of the neighbour galaxy. 
\textit{Lower panel}: same as upper panel having segregated the target
galaxies according to the morphology of their nearest neighbour. In this
case we do not present the dispersion on each bin for clarity}.
\end{figure}

In Cervantes-Sodi, Hernandez \& Park (2010), we found a weak but statistically
significant effect of galaxy-galaxy interactions on the $\lambda$ value of
late-type galaxies. These appear as soon as the galaxies cross into their
virial radii, leading to a gradual decrease in the values of $\lambda$, especially
when the neighbour galaxy is a spiral galaxy. For the galaxies in our sample, we
also test the influence of the small-scale environment. Fig 4 upper panel shows
$\lambda_{d}$ as a function of the distance to the nearest neighbour galaxy
normalized by the virial radius of the nearest neighbour as computed using
equation~\ref{Rvir}. In this case, we notice a decrease of the spin only for
the most inner bin, but given the small number of galaxies involved, the result
is also compatible with no dependence on the distance to the nearest neighbour galaxy.
For SDSS galaxies we previous noticed (see Fig 4 in Cervantes-Sodi et al 2010)
that the decrease of the spin depends on the morphology of the neighbour galaxy.
With this in mind we segregated the galaxies in our high-$z$ sample according
to the morphology of the neighbour galaxy to search for any difference on the
value of $\lambda_{d}$. Fig 4 lower panel shows the result, with only a significant
difference for the case of $d<0.1r_{vir}$, with a decrease of
the $\lambda_{d}$ value for the case of late-type neighbour and an increase
for the case of early-type neighbour, a
similar behaviour as the one observed with low redshift galaxies, but given
the small numbers involved it is difficult to reach a statistically meaningful
conclusion at the moment.

\section{Conclusions}

Using a simple dynamical model, we have obtained empirical
$\lambda$ distributions of high redshift galaxies in the range 
$0.4  \leq z \leq 1.2$ that are compatible with results
at low redshift. 

Analysing the angular momentum of the disc component, we did not
find traces of evolution as a function of redshift, and we were
able to reproduce the relationships of the spin with mass and
environment we found previously for local galaxies at all redshifts; showing a
marked dependence of $\lambda_{d}$ on the mass, with high mass galaxies showing typically
low $\lambda_{d}$ values, as well as lower spread when compared with low
mass galaxies; while there is little if any dependence of $\lambda_{d}$ on
environment, either large or small-scale environmental parameter. These results
lead us to conclude that the spin parameter of disc galaxies
suffers negligible evolution if any at all over the last 8 Gyr, which
is a result in good agreement with theoretical expectations
(Peirani, Mohayaee \& de Freitas Pacheco 2004; Kimm et al. 2011),
where most of the specific angular momentum is imprinted
at early stages (e.g. White 1984, Barnes \& Efsthathiou 1987, Catelan \& Theuns 1996,
Pichon et al. 2011), followed by mild evolution, with rare drastic
changes appearing only in the case of major mergers (e. g. Peirani et al.
2004, D'Onghia \& Navarro 2007). As shown by 
high-resolution cosmological simulations
(e. g. Brook et al. 2012, Governato 2007),
a relatively quite merging history also helps to form
realistic disc galaxies.
In addition, independent observational results suggest that most disc galaxies
are dynamically mature by $z \sim 1$ (Miller et al. 2012) in order to
reproduce the stellar mass TF relation that is already in place
at that redshift.

Finally, our results complement previous findings where different
groups have shown that the baryonic specific angular momenta of
disc galaxies at intermediate (Puech et al. 2007; Vergani et al. 2012)
and high redshift; nearly $z \sim 2$ (F\"orster Schreiber et al. 2006)
or even $z \sim 3$ (Nesvadba et al. 2006), are comparable to those of
local late-type galaxies. Our study thus bridges the gap between local samples and 
existing $z\sim 2 - 3$ results. The consistent conclusions across all redshift ranges
essentially rule out any significant dynamical evolution for large disk galaxies
over the last 8-10 Gyr.

\section*{Acknowledgments}

The authors acknowledge the thorough reading of the original
manuscript and comments by the referee, as helpful in reaching a
clearer and more complete final version.

This  work  is supported  by  NSFC   (no.   11173045),
Shanghai   Pujiang  Program (no. 11PJ1411600) and  the
CAS/SAFEA International Partnership Program for Creative
Research  Teams  (KJCX2-YW-T23).
XH acknowledges financial assistance 
from UNAM DGAPA grant IN103011.
CBP thanks Korea Institute for Advanced Study for providing
computing resources (KIAS Center for Advanced Computation
Linux Cluster System, QUEST).

\end{document}